\documentclass[12pt, fleqn]{article}
\usepackage{amsmath, amssymb}
\usepackage[dvips]{graphics,color}
\usepackage {graphicx}

\title{Point symmetries of 3D static plasma equilibrium systems: comparison and applications\\(Preprint)}
\author{Alexei F. Cheviakov\\
\small \emph{Department of Mathematics, University of British Columbia, Vancouver, V6T 1Z2 Canada}\\
\small\textbf{E-mail:} alexch@mast.ubc.ca}

\newtheorem{theorem}{Theorem}

\def\const{\hbox{\rm const}}

\def\grad{\mathop{\hbox{\rm grad}}}

\def\div{\mathop{\hbox{\rm div}}}
\def\curl{\mathop{\hbox{\rm curl}}}

\begin{document}
\maketitle
\begin{abstract}
Dynamic plasma equilibrium systems, both in isotropic and anisotropic framework, possess infinite-dimensional Lie groups of
point symmetries, which depend on solution topology and lead to construction of infinite families of new physical solutions.

By performing the complete classification, we show that in the static isotropic case no infinite point symmetries arise,
whereas static anisotropic plasma equilibria still possess a Lie group of symmetries depending on one free function defined
on the set of magnetic field lines. The finite form of the symmetries is found and used to obtain new exact solutions. We
demonstrate how anisotropic axially- and helically-symmetric equilibria are obtained using conventional Grad-Shafranov and
JFKO equations.

A recently developed multifunctional automated Maple-based software package for symmetry and conservation law analysis is
presented and used in this work.

\end{abstract}




\section{Introduction.}
\smallskip
The systems of isotropic Magnetohydrodynamics (MHD) and anisotropic Chew-Goldberger-Low (CGL) plasma equations, in
particular, their equilibrium reductions, are extensively used in physical applications, including the controlled
thermonuclear fusion research, geophysics and astrophysics (Earth magnetosphere, star formation, solar activity) and
laboratory and industrial applications \cite{fusion1,CGL,bisk}.

MHD and CGL systems, as well as their equilibrium versions, are essentially nonlinear systems of partial differential
equations, depending in general on three spatial coordinates $(x,y,z)$. Knowledge of physically meaningful exact solutions
and analytical properties of these systems (such as symmetries, conservation laws, stability criteria, etc.) is of great
importance for understanding the core properties of the underlying physical phenomena, for modeling, and for the development
of appropriate numerical methods.

Common ways of finding exact analytical solutions to such systems include: (1) PDE system reduction by a symmetry group
(self-similar or invariant solutions), (2) use of intrinsic symmetries of PDEs to generate new solutions from known ones,
and (3) use of transformations from solutions to other equations. The first approach has been successfully applied to the
static MHD system in cases of axial and helical symmetry (the well-known Grad-Shafranov \cite{gs1,gs2} and JFKO \cite{jfko}
equations), which resulted in several types of exact solutions (e.g. {\cite{ob_prl}}, {\cite{kadom}}-\cite{jets4}). A
different approach that makes use of equilibrium solution topology (general existence of 2D magnetic surfaces) has been
suggested in {\cite{afc_prl}}.

\bigskip
A symmetry of a system of DEs is any transformation of its solution manifold into itself. A symmetry thus transforms any
solution to another solution of the same system. Point symmetries are obtained by the general algorithmic Lie method of
group analysis \cite{olv}. It is equally applicable to algebraic and ordinary and partial differential equations. In a
similar algorithmic manner, contact and Lie-B\"{a}cklund symmetries (depending on derivatives) are found. Closely connected
is the potential symmetry method \cite{blum_kumei} for discovering nonlocal symmetries, and the notion of nonlocally-related
potential systems and subsystems \cite{BC1}.

\medskip
It is well-known that Lie group analysis method is capable of detecting not only several-parameter, but also
infinite-dimensional symmetry groups (depending on an arbitrary functions, see e.g. \cite{afc_pla, afc_kiev}.)

\bigskip
In this paper we perform Lie group analysis and the comparison of point symmetries of static isotropic (MHD) and anisotropic
(CGL) equations.

\medskip
Though the Lie procedure is straightforward and applicable to any ODE/PDE system with sufficiently smooth coefficients, it
requires extensive algebraic manipulation and the solution of (often large) overdetermined systems of linear PDEs. For many
contemporary models, \emph{especially those that do possess non-trivial symmetry structure,} such analysis presents a
significant computational challenge. For example, for static and dynamic MHD and CGL equilibrium systems, the determining
equations for the Lie point symmetries split into hundreds of linear equations.

We note that the recent derivation of important infinite Lie groups of symmetries of dynamic MHD {\cite{obsymm,obsymm3}} and
CGL {\cite{afc_ob}} equilibrium systems was done by special methods other than the Lie algorithm - the group analysis for
these systems has net been previously done due to computational difficulty. However later it was shown that Lie algorithm
indeed can detect those symmetries \cite{afc_pla, afc_kiev}.


\medskip
Modern analytical computation methods, based on Gr\"{o}bner bases {\cite{groeb1}}-{\cite{groeb4}} and characteristic sets
{\cite{charsets1}}-{\cite{charsets2}}, efficiently reduce large overdetermined systems of partial differential equations,
and in many cases facilitate the complete or partial group analysis. (For a review, see \cite{hereman_review}.)

One of the most powerful software packages for symmetry-related computations is a REDUCE-based set of programs by Thomas
Wolf \cite{t_wolf}. It includes CRACK, LIEPDE and CONLAWi routines for PDE system reduction, search for symmetries,
conservation laws and adjoint symmetries of DE systems.

\bigskip
In this work, we introduce a recently developed package \verb"GeM" for Maple {\cite{ac_package_www}} and use it for the
complete Lie group analysis of the static CGL and MHD systems. The package routines perform local (Lie, contact,
Lie-B\"{a}cklund) and nonlocal symmetry and conservation law analysis of ordinary and partial differential equations without
human intervention (Section \ref{a_pack}.) Corresponding determining systems are automatically generated and reduced, and in
most cases completely solved. The routines employ a special symbolic representation of the system under consideration, which
leads to the overall significant speedup of computations. This factor is of particular importance in the problems of
\emph{symmetry/conservation law classification} for systems with arbitrary constitutive function(s). The module routines can
handle DE systems of any order and number of equations, dependent and independent variables. To the best of the author's
knowledge, \verb"GeM" is the most functional and fast end-user-oriented package for Maple.

\medskip
The complete Lie group analysis and comparison of symmetries of static CGL and MHD systems reveals that when the anisotropy
factor is constant on magnetic field lines, the static CGL system possesses an \emph{infinite symmetry group}, involving an
arbitrary function constant on magnetic surfaces, similarly to the dynamical CGL equilibrium system (${\bf V}\neq 0$)
{\cite{afc_ob}}. On the other hand, unlike its dynamic version \cite{obsymm}, the static MHD system does not possess
infinite symmetries, but only affords scalings, translations and rotations (Section \ref{subs_sym_clas}.)

\medskip
The infinite symmetry group of the static CGL system deserves special attention. It depends on the solution topology, and
reveals an alternative PDE system representation that directly connects the static CGL system with the static MHD system,
establishing a "many-to-one, onto" solution correspondence (Sections \ref{subs_sym_clas}, \ref{subs_inf_prop}.) The latter
has important practical value. In particular, a direct way of construction of new CGL equilibria from known MHD ones
follows. For new CGL solutions obtained from a given MHD one, physical conditions and explicit stability criteria are
preserved, which makes them suitable for exact physical modelling. We also demonstrate how axially and helically symmetric
anisotropic static plasma equilibria can be constructed using the common form of Grad-Shafranov and JFKO equations (Section
\ref{subs_GS_CGL}.) (We remark that the equation derived in \cite{anis_gs} and referred to by authors as "The most general
form of the nonrelativistic Grad-Shafranov equation describing anisotropic pressure effects" is hardly usable, due to its
complicated form, and no nontrivial solutions to it are known.)

\medskip In Section \ref{subs_examp_vort}, we present an example of the application of the derived infinite symmetry group to
construction of a family of new exact static anisotropic (CGL) plasma equilibrium (an "anisotropic plasma vortex") in 3D
space from a known MHD equilibrium (which is at the same time a CGL equilibrium). Other types of exact static MHD solutions
known in literature may be used to produce families of CGL plasma equilibria in the similar manner.

\medskip Further remarks are presented in Section \ref{sec_further}.


\section{General Lie symmetry analysis of static MHD and CGL equations}\label{Gen_analys}

\subsection{The static MHD and CGL equations.}

The static MHD equilibrium system is obtained from the general MHD equations (e.g. {\cite{fusion1}}) in the time-independent
steady case ${\bf{V}}=0$ and has the form
\begin{equation}
{\rm{curl}}~{\bf{B}}\times{\bf{B}} = {\rm{grad}}~P , ~~{\rm{div}}~{\bf{B}}= 0, \label{eq_PEE}
\end{equation}
Here $\bf{B}$ is the vector of the magnetic field induction; $\rho$, plasma density; and $P$, plasma pressure.

The static time-independent equilibrium of anisotropic plasmas is a similar reduction of CGL equations {\cite{CGL,afc_ob}}
\begin{equation}
\left(1-\tau\right) {\rm{curl}}~{\bf{B}}\times{\bf{B}} = {\rm{grad}}~p_\perp +\tau~ {\rm{grad}}\frac{{{\bf{B}}^2 }}{2} +
\bf{B}(\bf{B}\cdot\rm{grad~}\tau), ~~{\rm{div}}~{\bf{B}}= 0.\label{eq_APEE}
\end{equation}
Here $\tau$ is the anisotropy factor,
\begin{equation}
\tau={\frac{p_\parallel-p_\perp}{\bf{B}^2}},~~~~\mathbb{P}_{ij}=p_\perp \delta_{ij} + \frac{p_\parallel-p_\perp}{B^2}B_i
B_j,~~~i,j=1,2,3,  \label{eq_tau_def}
\end{equation}
and the pressure $\mathbb{P}$ is a tensor with two independent parameters $p_\parallel, p_\perp$.

The anisotropic equilibrium system (\ref{eq_APEE}) is closed with an equation of state {\cite{afc_ob,afc_prl}}
\begin{equation}
\grad \tau \cdot {\bf{B}} = 0, \label{eq_tau_cond}
\end{equation}
which implies that the anisotropy factor $\tau$ is constant on magnetic field lines. Other possible choices include, for
instance, empiric relations (see {\cite{erkaev1}}.) (We remark that the relation (\ref{eq_tau_cond}) is compatible with
dynamic equations of state, such as "double-adiabatic" equations suggested in the original CGL paper {\cite{CGL}}, since
time derivatives identically vanish for all equilibria.)

\bigskip
It is known that in static MHD equilibria (\ref{eq_PEE}), the magnetic field ${\bf{B}}$ always lies on 2-dimensional
magnetic surfaces that span the plasma domain. The only possible exception is Beltrami-type configurations
$\curl{\bf{B}}=\alpha{\bf{B}}$, $\alpha=\const$. In a compact domain, magnetic surfaces are generally tori {\cite{kk}}.

In the general anisotropic equilibrium case, 2D magnetic surfaces are not necessarily present. However, as shown below, they
always exist in static anisotropic plasma configurations (\ref{eq_APEE})-(\ref{eq_tau_cond}).

\subsection{The Lie procedure.}\label{Sec_Lie_Proc}

For a general system of $l$ partial differential equations of order $p$,
\begin{equation}
\begin{array}{ll}
{\bf{E}}({\bf{x}}, {\bf{u}}, \mathop{\bf{u}}\limits_1,...,\mathop{\bf{u}}\limits_p)=0,\\
\displaystyle {\bf{E}} = (E^1,\dots,E^l),~{\bf{x}} =
(x^1,\dots,x^n)\in X, ~{\bf{u}} = (u^1,\dots,u^m) \in U, \\
\displaystyle \mathop{\bf{u}}\limits_k = \left( \frac{\partial^k u^j}{\partial x^{i_1}...\partial x^{i_k}}|~~ j=1,\dots,m
\right) \in U_k,~~k=1,\dots,p,
\end{array}
\label{e21}
\end{equation}
the set of all solutions is a manifold $\Omega$ in $(m+n)$ - dimensional space $X\times U$, which corresponds to a manifold
$\Omega^1$ in the prolonged (jet) space $X\times U\times U_1\times\dots\times U_p$ of dependent and independent variables
together with partial derivatives {\cite{olv}}.

In particular, for the systems (\ref{eq_PEE}) and (\ref{eq_APEE})-(\ref{eq_tau_cond}), the independent variables are
cartesian coordinates ${\bf{x}} = (x_1,x_2,x_3)$, and the dependent variables ${\bf{u}} = (B_1,B_2,B_3,P)$ and ${\bf{u}} =
(B_1,B_2,B_3,p_\perp,\tau)$ respectively ($n=3$, $m=4$ and $m=5$).

The Lie method of seeking Lie groups of transformations (symmetries)
\begin{equation}
\begin{array}{ll}
(x')^i  = f^i ({\bf{x}},{\bf{u}},\varepsilon) ~~(i=1,\dots,n),\\
(u')^j  = g^j ({\bf{x}},{\bf{u}},\varepsilon) ~~(j=1,\dots,m)
\end{array}
\label{e22}
\end{equation}
that map solutions of (\ref{e21}) into solutions consists in finding the Lie algebra of vector fields
\begin{equation}
\begin{array}{ll}
{\bf{v}}  =& \sum\limits_i {\xi^i({\bf{x}},{\bf{u}}) \frac{\partial }{{\partial x^i }}}  + \sum\limits_k
{\eta^k({\bf{x}},{\bf{u}}) \frac{\partial }{{\partial u^k }} } + \sum\limits_{i,\,k} {\xi _i^{(1) k}({\bf{x}}, {\bf{u}},
\mathop{\bf{u}}\limits_1) \frac{\partial }{{\partial u_i^k }}} \\
&+...+ \sum\limits_{i_1,...,i_p,~k} {\xi _{i_1,...,i_p}^{(p) k}({\bf{x}}, {\bf{u}},..., \mathop{\bf{u}}\limits_p)
\frac{\partial }{{\partial u_{i_1,...,i_p}^k }}}.
\end{array}\label{e23}
\end{equation}
tangent to the solution manifold $\Omega^1$ in the jet space. Here $\varepsilon$ is a group parameter, or a vector of group
parameters.

Here $\xi _{i_1,...,i_p}^{(p) k}$ are the coordinates of the prolonged tangent vector field corresponding to the derivatives
$u_{i_1,...,i_p}^k$; these are fully determined from ($\xi^i, \eta^k$) by differential relations. Thus tangent vector fields
(\ref{e23}) are isomorphic to infinitesimal operators
\begin{equation}
X  = \sum\limits_i {\xi^i({\bf{x}},{\bf{u}}) \frac{\partial }{{\partial x^i }}}  + \sum\limits_k {\eta^k({\bf{x}},{\bf{u}})
\frac{\partial }{{\partial u^k }}}.  \label{e26X}
\end{equation}

Operators (\ref{e26X}) are infinitesimal generators for the global Lie transformation group (\ref{e22}), which is
reconstructed explicitly by solving the initial value problem
\begin{equation}
\frac{{\partial f^i }}{{\partial \varepsilon}} = \xi ^i ({\bf{f}},{\bf{g}}), ~~ \frac{{\partial g^k }}{{\partial
\varepsilon}} = \eta ^k ({\bf{f}},{\bf{g}}), \label{e27_5}
\end{equation}
\begin{equation}
f^i (\varepsilon=0) = x^i, ~~ g^k (\varepsilon=0) = u^k.\nonumber
\end{equation}

\bigskip
To find all Lie group generators (\ref{e26X}) admissible by the original system (\ref{e21}), one needs to solve linear
determining equations
\begin{equation}
{\bf{v}} {\bf{E}}({\bf{x}}, {\bf{u}}, \mathop{\bf{u}}\limits_1,...,\mathop{\bf{u}}\limits_p)| _{ {\bf{E}}({\bf{x}},
{\bf{u}}, \mathop{\bf{u}}\limits_1,...,\mathop{\bf{u}}\limits_p) = {\rm{0}}}  = 0 \label{e28}
\end{equation}
on $m+n$ unknown functions $\xi^i, \eta^k$ that depend on $m+n$ variables $({\bf{x}}, {\bf{u}})$.

The overdetermined system on $\xi^i, \eta^k$ is obtained from ({\ref{e28}}) using the fact that $\xi^i, \eta^k$ do not
depend on derivatives $u^k_i$. Thus the system ({\ref{e28}}) splits into an $N \leq l(mn+1)$ linear partial differential
equations. For static MHD (CGL) equilibria, this leads to a system of 133 (253) linear PDEs on 7 (8) unknown functions
respectively. Methods of solution of such overdetermined systems are discussed in the following subsection.

\bigskip
Tangent vector field components $\xi^i, \eta^k$ of the symmetry generator (\ref{e23}) may be allowed to depend on
derivatives $\mathop{\bf{u}}\limits_1,...,\mathop{\bf{u}}\limits_p.$ Such symmetries are known as \emph{Lie-B\"{a}cklund
symmetries} {\cite{olv}}. Unlike Lie symmetries, they act on the solution manifold $\Omega_{\infty}$ in the
infinite-dimensional space $X\times U\times U_1\times\dots\times U_p\times\dots$ of all variables and derivatives. Due to
this fact, Lie-B\"{a}cklund symmetries can not be integrated to yield a global group representation similar to (\ref{e22}),
and can not be used to explicitly map solutions of differential equations into new solutions.

\subsection{"GeM" symbolic package for general Lie symmetry computation.}\label{a_pack}

For a system of several PDEs with several dependent and independent variables, overdetermined systems resulting from
determining equations can be large - consisting of dozens or hundreds of linear PDEs. The simplification and integration of
such systems, especially in the cases of nontrivial symmetry structure (e.g. infinite symmetries), often presents a
computational challenge. Analytical computation software is traditionally used to produce and reduce large linear PDE
systems that arise.

\medskip
Algorithms for reduction overdetermined systems based on Gr\"{o}bner bases and characteristic sets techniques have been
successfully implemented \cite{hereman_review}. However the author had difficulty finding an "end-user" package that does
not require human intervention in computation, has sufficient functionality (i.e. is capable of computing Lie and
Lie-B\"{a}cklund symmetries of arbitrary ODE/PDE systems of reasonable order, performing symmetry classification with
respect to constitutive functions of the system, doing conservation law analysis, etc.) in conceivable time.

Available general Lie group analysis software packages, such as \verb"Desolv" by Vu and Carminati, partly have desired
functionality, including no user intervention in symmetry computation and output in the form of a set of symmetry generators
({\ref{e23}}); however, for the systems with several spatial and dependent variables, such as static MHD equilibrium system
(\ref{eq_APEE}), the analysis can not be performed with usual computation resources in finite time (though the Lie symmetry
group finite-dimensional).

\bigskip
A Waterloo Maple - based package \verb"GeM" ("General Module") has been recently developed by the author. The package
routines are capable of finding all (Lie, contact and Lie-B\"{a}cklund) local symmetries and prescribed classes of
conservation laws for any ODE/PDE system without significant limitations on DE order and number of variables, and without
human intervention.

The routines of the module allow the analysis of ODE/PDE systems containing arbitrary functions. \emph{Classification} and
isolation of functions for which additional symmetries / conservation laws occur is automatically accomplished. Such
classification problems naturally occur in the analysis of DE systems that involve constitutive functions. The symmetry
classification may lead to the linearization, discovery of new conservation laws and symmetries in particular cases (for
example, see \cite{BC1,afc_pla}.)

The "GeM" package \cite{ac_package_www} employs a special representation of the system under consideration and the
determining equations: all dependent variables and derivatives are treated as Maple \emph{symbols}, rather than functions or
expressions. This significantly speeds up the computation of the tangent vector field components corresponding to
derivatives (up to any order) and the production of the (overdetermined) system of determining equations on the unknown
tangent vector field coordinates. The latter is reduced by \verb"Rif" \cite{RifDesc} package routines (which is also done
faster with derivatives being symbols, rather than expressions.)

\medskip
The reduced system can, in most cases, be automatically and completely integrated by Maple's internal \verb"PDSolve"
routine. However (in rare cases) the routine is known to miss solutions, therefore it is recommended that the solution of
the reduced determining system (for symmetry or conservation law analysis) is verified by hand in each case.

\medskip
We would like to underline that in the study or classification of symmetries and conservation laws, the most important and
time-consuming task is indeed the simplification (and in the case of a classification problem, the case-splitting) of the
overdetermined PDE system of determining equations. This task is automatically done by \verb"GeM" routines. The consecutive
task of solving the reduced system is in most cases much simpler, and usually can be done and verified by hand in reasonable
time.

\bigskip
A brief description of symmetry analysis procedure using the \verb"GeM" package is given in Appendix A.

\section{Comparison of point symmetries of static MHD and CGL equilibrium systems}\label{Point_Transfs}

In the current section we present the complete Lie point symmetry classification of static MHD and CGL systems
(\ref{eq_PEE}), (\ref{eq_APEE})-(\ref{eq_tau_cond}), and demonstrate that the static CGL system possesses an
infinite-dimensional symmetry group (depending on an arbitrary function), whereas the static CGL system does not.

Global transformation groups corresponding to the discovered symmetry generators are given. Special attention is devoted to
an infinite-dimensional symmetry group of the CGL static equilibrium system, that appears to be related to the
infinite-dimensional symmetry group of dynamic CGL equilibrium system \cite{afc_ob}. (This is not the case for the static
MHD system - it does not possess infinite-dimensional symmetry groups, though its dynamic version does \cite{obsymm}.)

The infinite symmetries of the static CGL system (\ref{eq_APEE})-(\ref{eq_tau_cond}) directly relate it to static MHD
equations, which is important for applications. Namely, it provides a way of construction of exact anisotropic plasma
equilibria from known static MHD solutions, allow simple formulation of famous Grad-Shafranov and JFKO equations for
anisotropic plasmas, and more (see Section \ref{Examp_sec}.)

(The "GeM" package described above has been used for computations.)

\subsection{Symmetry classification}\label{subs_sym_clas}

\bigskip
\begin{theorem}[Point symmetries of static CGL and MHD equations] \label{th_01}

(i)The basis of Lie algebra of point symmetry generators of the general static CGL equilibrium system {(\ref{eq_APEE})}
consists of the operators
\begin{equation}
X_{trans}  = \sum\limits_{k=1}^3 K_i \frac{\partial }{{\partial x_i }} + K_4\frac{\partial }{{\partial p_\perp}};\nonumber
\end{equation}
\begin{eqnarray*}
X_{rot}  = (c B_2+d B_3)\frac{\partial }{\partial B_1}+(-cB_1-bB_3)\frac{\partial }{\partial B_2}+(-dB_1+bB_2)\frac{\partial
}{\partial B_3}\\
 +(cx_2+dx_3)\frac{\partial}{\partial x_1}+(-cx_1-bx_3)\frac{\partial}{\partial x_2}+(-dx_1+bx_2)\frac{\partial}{\partial
x_3};
\end{eqnarray*}
\begin{equation}
X_{scal}^{(1)} = \sum\limits_{k=1}^3 x_i \frac{\partial }{{\partial x_i }};~~~X_{scal}^{(2)} = \left( \sum\limits_{k=1}^3
B_i \frac{\partial }{{\partial B_i }} + 2 p_\perp \frac{\partial }{{\partial p_\perp}}\right); \label{eq_thrm_pee_gen}
\end{equation}
\begin{equation}
X_{scal}^{(3)} = \left(\left(p_\perp+\frac{\bf{B}^2}{2}\right) \frac{\partial }{\partial p_\perp} -
(1-\tau)\frac{\partial}{\partial \tau}\right), \nonumber
\end{equation}
corresponding to translations, rotations and scalings ($b,c,d,t,s,w, K_1,...,K_4$ are real parameters.)

\bigskip
(ii) The static anisotropic CGL plasma equilibrium system (\ref{eq_APEE}) with equation of the state (\ref{eq_tau_cond})
affords an additional infinite-dimensional Lie group of symmetries generated by
\begin{equation}
X_{\infty}  = F\left(\tau, p_\perp + \tau \frac{{\bf{B}}^2}{2}\right) \cdot \left( \sum\limits_{k=1}^3 B_i \frac{\partial
}{{\partial B_i }} + 2(1-\tau)\frac{\partial }{{\partial \tau }} - {\bf{B}}^2 \frac{\partial }{{\partial p_\perp}}\right),
\label{eq_inf_symm_gen}
\end{equation}
where $F$ is a sufficiently smooth function of its arguments.

\bigskip
(iii) The Lie algebra of Lie point symmetry generators of static MHD equilibrium system {(\ref{eq_PEE})} is spanned by the
infinitesimal $X_{trans}, X_{rot}, X_{scal}^{(1)}, X_{scal}^{(2)}$ (with $p_\perp$ replaced by $P$).

\end{theorem}

\bigskip \noindent \textbf{Proof sequence.}

\smallskip\noindent(i). The system of determining equations for the static CGL equilibrium system {(\ref{eq_APEE})} consists of four equations.
After setting all coefficients at derivatives to zero, an overdetermined system of 253 equations obtained (using the
\verb"acgen_get_split_sys()" routine.) After simplification with \verb"rifsimp()" routine, the system is reduced to 64
simpler equations (most of which are conditions of the form $\partial \xi^k/ \partial x^m =0$, which are readily integrated
by hand to produce generators {(\ref{eq_thrm_pee_gen})}.

\smallskip\noindent(ii). For the CGL system (\ref{eq_APEE}) with the equation of state (\ref{eq_tau_cond}), the picture is more complicated. Five determining equations split
into an overdetermined system of 199 linear equations. It is reduced to 49 simpler equations, which show that $\eta^k$ are
independent of the spatial variables, whereas $\xi^k$ depend on $x_1,x_2,x_3$ linearly, providing rotations and scalings.
After factoring out the latter, 15 equations are left, from which tangent vector field coordinates of
{(\ref{eq_inf_symm_gen})} are found by straightforward integration.

\smallskip\noindent(iii). For the static MHD system {(\ref{eq_PEE})}, four determining equations are split into an overdetermined system of 133 equations
After simplification with \verb"rifsimp()", the system is reduced to 49 simpler equations (most of which in the form of a
partial derivative equal to zero), which are integrated by hand and give rise to operators  $X_{trans}, X_{rot},
X_{scal}^{(1)}, X_{scal}^{(2)}$.

This completes the proof of the theorem. $\square$

\bigskip \noindent \textbf{Finite form of point transformations.}

Using the reconstruction formula (\ref{e27_5}), the global transformation groups corresponding to symmetry generators
(\ref{eq_thrm_pee_gen}) are found:

\bigskip \noindent \textbf{1.}$X_{trans}$: translations
\begin{equation}
{x_i}'=x_i + K_i \varepsilon, ~~i=1,2,3;~~~ p_\perp'=p_\perp + K_4 a,~~{\bf B}'={\bf B},~~{\tau}'=\tau;\nonumber
\end{equation}

\bigskip \noindent \textbf{2.}$X_{rot}$: $SO_3$ group of 3D space rotations, parameterized, for example, by Euler angles $\phi,\theta,\psi$:
\begin{equation}
{\bf x}'=BCD {\bf x};~~~{\bf B}'({\bf x}')=BCD {\bf B}({\bf x});~~~p_\perp'({\bf x}')=p_\perp({\bf x}),~~{\tau}'({\bf
x}')=\tau({\bf x}), \nonumber
\end{equation}
\begin{equation}\nonumber
B=\left(\begin{array}{ccc}\cos \phi & \sin \phi & 0\\-\sin \phi & \cos \phi & 0\\0&0&1\end{array}\right);~~
C=\left(\begin{array}{ccc}1&0&0\\0 &\cos \theta & \sin \theta \\0&-\sin \theta & \cos \theta \end{array}\right);~~
B=\left(\begin{array}{ccc}\cos \psi & \sin \psi & 0\\-\sin \psi & \cos \psi & 0\\0&0&1\end{array}\right).
\end{equation}

\bigskip \noindent \textbf{3.}$X_{scal}^{(1)},X_{scal}^{(2)}$: to independent scalings of coordinates and dependent variables
\begin{equation}
{x_i}'=t x_i,~~{B_i}'=s B_i,~~ p_\perp'=2sp_\perp,~~{\tau}'=\tau, ~~~i=1,2,3;~~t,s\in \mathbb{R}. \nonumber
\end{equation}

\bigskip \noindent \textbf{4.}$X_{scal}^{(3)}$: a group of scalings
\begin{equation}
{x_i}'=x_i,~~{B_i}'=B_i,~~i=1,2,3;~~~
\left(p_\perp+\frac{\bf{B}^2}{2}\right)'=C\left(p_\perp+\frac{\bf{B}^2}{2}\right),~~(1-{\tau})'=C(1-\tau);~~C\in \mathbb{R}.
\nonumber
\end{equation}

\bigskip

\bigskip
\begin{theorem}[The infinite-dimensional symmetry group corresponding to $X_{\infty}$]\label{theorem_2}

The static anisotropic CGL equilibrium system (\ref{eq_APEE})-(\ref{eq_tau_cond}) possesses an infinite-dimensional set of
point transformations of solutions into solutions given by formulas
\begin{equation}
{\bf{B}}_1  = {\bf{B}} M(\Psi),~~\tau_1=1-(1-\tau)M^{-2}(\Psi), \label{eq_coroll1_1}
\end{equation}
\begin{equation}
p_{\perp 1}= p_{\perp} + \frac{{\bf{B}}^2-{\bf{B}}_1^2}{2}, \label{eq_coroll1_2}
\end{equation}
These transformations form a Lie group generated by the operator $X_{\infty}$ (\ref{eq_inf_symm_gen}). Here $\Psi=\Psi({\bf
r})$, ${\bf r}\in \mathbb{R}^3$, is a function constant on magnetic field lines (i.e. on magnetic surfaces when they exist.)

Moreover, the static anisotropic CGL equilibrium system (\ref{eq_APEE})-(\ref{eq_tau_cond}) can be written in the form
\begin{equation}
\curl\left(\sqrt{1-\tau}{\bf{B}}\right)  \times \left(\sqrt{1-\tau}{\bf{B}}\right) = \grad\left(p_\perp + \tau
\frac{{\bf{B}}^2}{2}\right),\label{eq_APEE2_1}
\end{equation}
\begin{equation}
\div~{\bf{B}}= 0,~~~{\bf{B}}\cdot\grad\tau=0, \label{eq_APEE2_2}
\end{equation}
which implies that $p_\perp + \tau {{\bf{B}}^2}/{2}$ is constant on magnetic field lines:
\begin{equation}
{\bf{B}}\cdot\grad\left(p_\perp + \tau \frac{{\bf{B}}^2}{2}\right)=0, \label{eq_APEE2_3}
\end{equation}
\end{theorem}

\bigskip
The proof of this theorem is given in Appendix B.

\bigskip \noindent \textbf{Remark 1.} According to the definition (\ref{eq_tau_def}) of the anisotropy factor $\tau$, the
parallel pressure component $p_{\parallel}$ transforms as follows:
\begin{equation}
p_{\parallel 1}= p_{\perp 1} +
{\bf{B}}_1^2\left(1-\left(1-\frac{p_{\parallel}-p_{\perp}}{{\bf{B}}^2}\right)M^{-2}(\Psi)\right). \label{eq_coroll1_3}
\end{equation}

\bigskip \noindent \textbf{Remark 2.}
The form of the arbitrary function in the symmetry generator $X_{\infty}$ uncovers the fact that the quantity $p_\perp +
\tau {{\bf{B}}^2}/{2}$ , together with the anisotropy parameter $\tau$, is constant on magnetic field lines, and leads to an
alternative representation (\ref{eq_APEE2_1})-(\ref{eq_APEE2_2}) of the static CGL equilibrium system
(\ref{eq_APEE})-(\ref{eq_tau_cond}).

\subsection{Properties of the infinite symmetries (\ref{eq_coroll1_1})-(\ref{eq_coroll1_2}) of the static CGL
system}\label{subs_inf_prop}

As remarked in the Introduction, both MHD and CGL \emph{dynamic} equilibrium systems (${\bf{V}}\neq 0$) possess
infinite-dimensional symmetries. However it is not so in the \emph{static} case ${\bf{V}}= 0$: the MHD system (\ref{eq_PEE})
only admits scalings, rotations and translations, whereas the static anisotropic (CGL) plasma equilibrium system
(\ref{eq_APEE})-(\ref{eq_tau_cond}) still does possess an infinite-dimensional set of symmetries.

\bigskip The infinite symmetries (\ref{eq_coroll1_1})-(\ref{eq_coroll1_2}), (\ref{eq_coroll1_3}) of the static CGL system
constitute a subgroup of the infinite-dimensional symmetry group $G$ of dynamic CGL equilibrium system found in
\cite{afc_ob}, and share many of their properties. Most important properties of transformations and new solutions
constructed using them are listed below.

\bigskip \noindent \textbf{Structure of the arbitrary function.} The transformations (\ref{eq_coroll1_1})-(\ref{eq_coroll1_2}), (\ref{eq_coroll1_3})
depend on the topology of the initial solution they are applied to, namely, on the set of magnetic field lines
$\Psi=\const$. Here $\Psi$ is  an arbitrary sufficiently smooth real-valued function constant on magnetic field lines of the
original static CGL plasma equilibrium $\{{\bf{B}},p_{\perp},p_{\parallel}\}$. In the following topologies, the domain of
the function $\Psi=\Psi(\bf{r})$ is evident:

\smallskip \emph{(a).} The lines of magnetic field ${\bf{B}}$ are closed loops or go to infinity. Then the function
$\Psi(\bf{r})$ has to be constant on these lines.

\smallskip \emph{(b).} The magnetic field lines are dense on 2D magnetic surfaces spanning the plasma domain $\mathcal{D}$. Then the function $\Psi(\bf{r})$ has
a constant value in on each magnetic surface, and is generally a function defined on a \emph{cellular complex} (a
combination of 1-dimensional and 2-dimensional sets) determined by the topology of the initial solution
$\{{\bf{B}},p_{\perp},p_{\parallel}\}$.

\smallskip \emph{(c).} The magnetic field lines are dense in some 3D domain $\mathcal{A}$. Then the function $\Psi(\bf{r})$ is constant in $\mathcal{A}$.

\bigskip \noindent \textbf{Topology of new solutions. Boundary conditions. Physical properties.}
New solutions $\{{\bf{B}}_1,p_{\perp 1},p_{\parallel 1}\}$ generated by the transformations
(\ref{eq_coroll1_1})-(\ref{eq_coroll1_3}) do not change the direction of ${\bf{B}}$, and thus retain the magnetic field
lines of the original plasma configuration $\{{\bf{B}},p_{\perp},p_{\parallel}\}$. Therefore usual boundary conditions for
the plasma equilibrium of the type ${\bf{n}}\cdot{{\bf{B}}}|_{\partial\mathcal{D}} = 0$ (${\bf{n}}$ is a normal to the
surface of the plasma domain $\mathcal{D}$) are preserved.

If the arbitrary function $M(\Psi)$ is separated from zero, then the transformed static anisotropic solutions retain the
boundedness of the original solution; the same is true about the magnetic energy ${{\bf{B}}^2}/2.$ In models with infinite
domains, the free function must be chosen so that the new anisotropic solution has proper asymptotics at
$|{\bf{r}}|\rightarrow \infty$.

\bigskip \noindent \textbf{Stability of new solutions.} No general stability criterion is available
for MHD or CGL equilibria. However, for particular types of plasma instabilities, explicit criteria have been found. In
particular, under the assumption of double-adiabatic behaviour of plasma {\cite{CGL}} the criterium for the fire-hose
instability is {\cite{cgl_instab}}
\begin{equation}
p_{\parallel} - p_{\perp} > {\bf{B}}^2, \label{e_instab1}
\end{equation}
(or, equivalently, $\tau>1$), and for the mirror instability --
\begin{equation}
p_{\perp} \left(\frac{p_{\perp}}{6 p_{\parallel}} - 1\right)
> \frac{{\bf{B}}^2}{2}. \label{e_instab2}
\end{equation}

If the initial solution $\{{\bf{B}},p_{\perp},p_{\parallel}\}$ is stable with respect to the fire-hose instability
($\tau<1$), then according to (\ref{eq_coroll1_1})
\begin{equation}
1-\tau_1=(1-\tau)/M^{2}(\Psi)>0, \nonumber
\end{equation}
and $\tau_1<1$, hence \emph{transformed solutions retain fire-hose stability/instability of the original ones}.

\bigskip
It can be shown that for every initial configuration $\{{\bf{B}},p_{\perp},p_{\parallel}\}$, there exists a nonempty range
of values which $M^{2}(\Psi)$ may take so that \emph{the mirror instability also does not occur}. The proof is parallel to
that in Section 2 of {\cite{afc_ob}}.

\bigskip \noindent \textbf{Group structure.}
We consider the set $G_C$ of all transformations (\ref{eq_coroll1_1}), (\ref{eq_coroll1_2}) with smooth $M(\Psi)$. Each
transformation (\ref{eq_coroll1_1}), (\ref{eq_coroll1_2}) is uniquely defined by a pair $\{\alpha, H(\Psi)\}$ as follows:
\begin{equation}
M\left(\Psi\right)=\alpha \exp\left(H\left(\Psi\right)\right), ~~\alpha = \pm1. \nonumber
\end{equation}

The composition of two transformations $(\alpha, H(\Psi))$ and $(\beta, K(\Psi))$ is equivalent to a commutative group
multiplication
\begin{equation}
(\alpha,H)\cdot(\beta, K) = (\alpha\beta, H+K),\nonumber
\end{equation}
and the inverse $(\alpha, H)^{-1} =$ $(\alpha, -H)$. The group unity is $e=(1, 0)$. Thus the set $G_C$ is an abelian group
\begin{equation}
G_C = A_{\Psi} \oplus Z_2.\label{new_group_st}
\end{equation}
It has two connected components; $A_{\Psi}$ is the additive abelian group of smooth functions in $\mathbb{R}^3$ that are
constant on magnetic field lines of a given static CGL configuration.


\section{Applications of infinite-dimensional transformations of the static CGL system}\label{Examp_sec}

\subsection{The Grad-Shafranov and JFKO equations for anisotropic plasmas.}\label{subs_GS_CGL}

In 1958, Grad and Rubin {\cite{gs1}} and Shafranov {\cite{gs2}} have independently shown that the system (\ref{eq_PEE}) of
static isotropic MHD equilibrium equations having axial symmetry (independent of the polar angle) is equivalent to one
scalar equation, called \textit{Grad-Shafranov equation}:
\begin{equation}
\psi_{rr} - \frac{\psi_r }{r} + \psi_{zz} + I(\psi){I}'(\psi) = - r^2 {P}'(\psi). \label{eq_GS}
\end{equation}

The magnetic field \textbf{B} and pressure $P$ have the form
\begin{equation}
{\rm {\bf B}} = \frac{\psi_z }{r}{\rm {\bf e}}_r + \frac{I(\psi)}{r}{\rm {\bf e}}_\varphi - \frac{\psi_r }{r}{\rm {\bf
e}}_z,~~~P=P(\psi)\nonumber
\end{equation}
($r$, $\varphi $, $z)$ are cylindrical coordinates. Surfaces $\psi=\const$ are the magnetic surfaces.

\bigskip
The JFKO equation is another reduction of the system of static MHD equilibrium equations (\ref{eq_PEE}), which describes
helically symmetric plasma equilibrium configurations, i.e. configurations invariant with respect to the helical
transformations
\begin{equation}
z \to z + \gamma h,\,\,\,\varphi \to \varphi + h,\,\,\,r \to r,\label{eq_helical_coord}
\end{equation}
\noindent where ($r$, $\varphi $, $z)$ are cylindrical coordinates. Solutions to the JFKO equation therefore depend only on
($r$, $u)$, where $u=z-\gamma \varphi $. The equation was derived by Johnson, Oberman, Kruskal and Frieman {\cite{jfko}} in
1958, and may be written in the form
\begin{equation}
\frac{\psi _{uu} }{r^2} + \frac{1}{r}\frac{\partial }{\partial r}\left( {\frac{r}{r^2 + \gamma ^2}\psi _r } \right) +
\frac{I(\psi){I}'(\psi)}{r^2 + \gamma ^2} + \frac{2\gamma I(\psi )}{\left( {r^2 + \gamma ^2} \right)^2} = - {P}'(\psi ).
\label{eq_jfko}
\end{equation}

The helically symmetric magnetic field is
\begin{equation}
{\bf B}_h = \frac{\psi_u}{r}{\bf{e}}_r + B_1{\bf {e}}_z + B_2{\bf {e}}_\phi, \quad B_1 = \frac{\gamma I(\psi) - r\psi_r}{r^2
+ \gamma^2}, \quad B_2 = \frac{r I(\psi) + \gamma \psi_r}{r^2 + \gamma^2}, \nonumber
\end{equation}
($r$, $\varphi $, $z)$ are cylindrical coordinates. $I(\psi)$ and pressure $P(\psi)$ are arbitrary functions. As in the
Grad-Shafranov reduction, the magnetic surfaces here are enumerated by the values of the flux function
$\psi=\psi(r,u)=\const$.

\bigskip
The direct analogy between the system (\ref{eq_PEE}) of static isotropic (MHD) plasma equilibrium equations and the
alternative representation (\ref{eq_APEE2_1})-(\ref{eq_APEE2_2}) of the static anisotropic (CGL) equilibrium system leads to
the following theorem.

\begin{theorem}[Axially and helically symmetric static anisotropic plasma equilibria]
(i). All axially symmetric solutions of the system of anisotropic (CGL) plasma equilibrium equations
(\ref{eq_APEE})-(\ref{eq_tau_cond}) in 3D space are found from solutions of a Grad-Shafranov equation
\begin{equation}
\Psi_{rr} - \frac{\Psi_r }{r} + \Psi_{zz} + J(\Psi){J}'(\Psi) = - r^2 N'(\Psi), \label{eq_AGS1}
\end{equation}
where $\Psi(r,z)$ is a flux function, $J(\Psi)$ and $\displaystyle N(\Psi)$ 
 are some functions of $\Psi(r,z)$, and have the form
 \begin{equation}\label{eq_AGS2}
\begin{array}{ll}
\tau=\tau(\Psi)<1,\\
\displaystyle {\bf B} = \frac{1 }{\sqrt{1-\tau}} \left(\frac{\Psi_z }{r}{\rm {\bf e}}_r + \frac{J(\Psi)}{r}{\rm {\bf
e}}_\varphi - \frac{\Psi_r }{r}{\rm {\bf e}}_z\right),\\
\displaystyle p_\perp=N(\Psi)-\frac{\tau {\bf B}^2}{2},~~~ \displaystyle p_{\parallel}= N(\Psi)+\frac{\tau {\bf B}^2}{2}.
\end{array}
\end{equation}

Conversely, any solution to the Grad-Shafranov equation (\ref{eq_AGS1}) corresponds to a family of axially symmetric
anisotropic plasma equilibria determined by (\ref{eq_AGS2}).

(ii). All helically symmetric solution of the system of static anisotropic (CGL) plasma equilibrium equations
(\ref{eq_APEE})-(\ref{eq_tau_cond}) in 3D space correspond to solutions of the JFKO equation
\begin{equation}\label{eq_AJFKO1}
\frac{\Psi _{uu} }{r^2} + \frac{1}{r}\frac{\partial }{\partial r}\left( {\frac{r}{r^2 + \gamma ^2}\Psi _r } \right) +
\frac{J(\Psi){J}'(\Psi)}{r^2 + \gamma ^2} + \frac{2\gamma J(\Psi )}{\left( {r^2 + \gamma ^2} \right)^2} = - {L}'(\psi ).
\end{equation}
where $\Psi(r,u)$ is a flux function, $J(\Psi)$ and $\displaystyle L(\Psi) $
are some functions of $\Psi(r,u)$, and the helically ymmetric static CGL plasma equilibrium is given by
\begin{equation}\label{eq_AJFKO2}
\begin{array}{ll}
\tau=\tau(\Psi)<1,\\
\displaystyle {\bf B} =  \frac{1 }{\sqrt{1-\tau}} \left(\frac{\Psi_u}{r}{\bf{e}}_r + B_1{\bf {e}}_z + B_2{\bf
{e}}_\phi\right), \quad B_1 = \frac{\gamma J(\Psi)
- r\Psi_r}{r^2+ \gamma^2}, \quad B_2 = \frac{r J(\Psi) + \gamma \Psi_r}{r^2 + \gamma^2},\\
\displaystyle p_\perp=L(\Psi)-\frac{\tau {\bf B}^2}{2},~~~ \displaystyle p_{\parallel}= L(\Psi)+\frac{\tau {\bf B}^2}{2}.
\end{array}
\end{equation}
Conversely, any solution to the JFKO equation (\ref{eq_AJFKO1}) corresponds to a family of helically symmetric anisotropic
plasma equilibria determined by (\ref{eq_AJFKO2}).

\end{theorem}

The theorem statement directly follows from the equivalence of the equations (\ref{eq_APEE2_1})-(\ref{eq_APEE2_2}.1) and the
system (\ref{eq_PEE}). The equation (\ref{eq_APEE2_2}.2) provides that the anisotropy factor $\tau=\tau(\Psi)$ is constant
on axially- (helically-) symmetric magnetic surfaces $\Psi=\const$.

\bigskip \noindent \textbf{Remark.} Several types of solutions to Grad-Shafranov and JFKO equations are available in literature. Examples of solutions
with physical behaviour can be found in Refs. \cite{ob_prl, jets4, ball2}.


\subsection{Example of an exact solution: an anisotropic plasma vortex.}\label{subs_examp_vort}

Below we use the infinite point symmetries (\ref{eq_coroll1_1})-(\ref{eq_coroll1_2}), (\ref{eq_coroll1_3}) of static CGL
equations to obtain a vortex-like exact static anisotropic plasma configuration from a known isotropic one.

\bigskip
In \cite{ball4}, Bobnev derived a localized vortex-like solution to the static isotropic MHD equilibrium system
(\ref{eq_PEE}) in 3D space. The solution is presented in spherical coordinates $(\rho, \theta, \phi)$ and is axially
symmetric, i.e. independent of the polar variable $\phi$. \footnote{Though Bobnev's solution was found by an \emph{ad hoc}
method without using Grad-Shafranov equation, it indeed corresponds to some solution of the Grad-Shafranov equation
(\ref{eq_GS}). However, it is not possible to write the Grad-Shafranov flux function $\psi$ for this solution in the closed
form, due to the use of spherical coordinates.}

The solution has the form
\begin{equation} \label{bob_sol1}
{\bf B}_v={\bf e}_{\rho}V(\rho)\cos(\theta)+{\bf e}_{\theta}W(\rho)\sin(\theta)+{\bf
e}_{\phi}U(\rho)\sin(\theta),~~~P_v=P_0-p(\rho)\sin^2(\theta),
\end{equation}
where
\begin{equation} \label{bob_sol2}
\begin{array}{ll}
B_0, P_0 = \const,\\

U(\rho) = \lambda_n \rho V(\rho),~~p(\rho) = \gamma \rho^2 V(\rho),~~W(\rho) = -V(\rho)-\rho V'(\rho)/2,\\

\displaystyle V(\rho)=B_0 \frac{V_0(2\lambda_n\rho)-V_0(2\lambda_n R)}{1-V_0(2\lambda_n R)},~~\gamma=B_0
\frac{V_0(2\lambda_n R)}{1-V_0(2\lambda_n R)}=\const,\\

\displaystyle V_0(x)=3\left(\frac{\sin x}{x^3}-\frac{\cos x}{x^2}\right).
\end{array}
\end{equation}

$\lambda_n,~ n=1,2,...$ is one of countable set of solutions of the equation
\begin{equation}
(3-4 R^2 \lambda_n^2) \sin(2R \lambda_n)-6 R \lambda_n \cos(2R\lambda_n)=0,~~R=\const.
\end{equation}
(In particular, $R \lambda_1\approx 2.882, R \lambda_2\approx 4.548, R \lambda_3 \approx 6.161$.) Bobnev's solution
satisfies boundary conditions
\begin{equation}
{\bf B}|_{\rho=R}=0,~~P|_{\rho=R}=P_0,~~{\bf B}|_{\rho=0}=B_0 {\bf e}_z,
\end{equation}
i.e. the region where the magnetic field is nonzero is a sphere of radius $R$. The magnetic surfaces $\Psi=\const$ inside
the sphere are families of inscribed tori of non-circular section, separated by separatrices; the number and mutual position
of the families depends on the choice of value of $R, \lambda_n$.

\bigskip
We use the the point symmetries (\ref{eq_coroll1_1}),(\ref{eq_coroll1_2}),(\ref{eq_coroll1_3}), starting with the initial
solution $({\bf{B}},\tau, p_{\perp}) \equiv ({\bf{B}}_v,0, P_v)$ given by (\ref{bob_sol1}), (\ref{bob_sol2}), corresponding
to $R=1, \lambda_3 \approx 6.161$. For this MHD equilibrium, the profile of levels of constant magnetic energy ${\bf
B}_v^2/2=\const$ and pressure $P_v=\const$ are shown on Fig. 1 (left and right, respectively.) The set of magnetic surfaces
$\Psi=P/P_{max}$ coincides with levels of constant pressure $P_v=\const$.

\bigskip
As an example, in transformation folmulas (\ref{eq_coroll1_1}),(\ref{eq_coroll1_2}),(\ref{eq_coroll1_3}), we choose the
arbitrary function $M(\Psi)=1+\Psi \sin(\Psi)$, which is separated from zero (since $|\Psi|\leq 1$) and constant on the
magnetic surfaces of the original static MHD configuration.

\bigskip
Fig. 2 shows the surfaces of constant level of anisotropic pressure components $P,~p_{\perp}~,p_{\parallel}$, and their
profile along the radius of the vortex in the direction perpendicular to $z$.

Fig. 3 illustrates the profiles of magnetic energy densities ${\bf{B}}_v^2/2$ and ${\bf{B}}_1^2/2$ for isotropic and
anisotropic plasma vortex (left), and the profiles of anisotropic pressure components $p_{\perp}~,p_{\parallel}$, in the
radial direction (perpendicular to the axis of symmetry $z$.)


\begin{figure}[htbp]
\centerline{
\includegraphics[width=2.5in,height=2.5in]{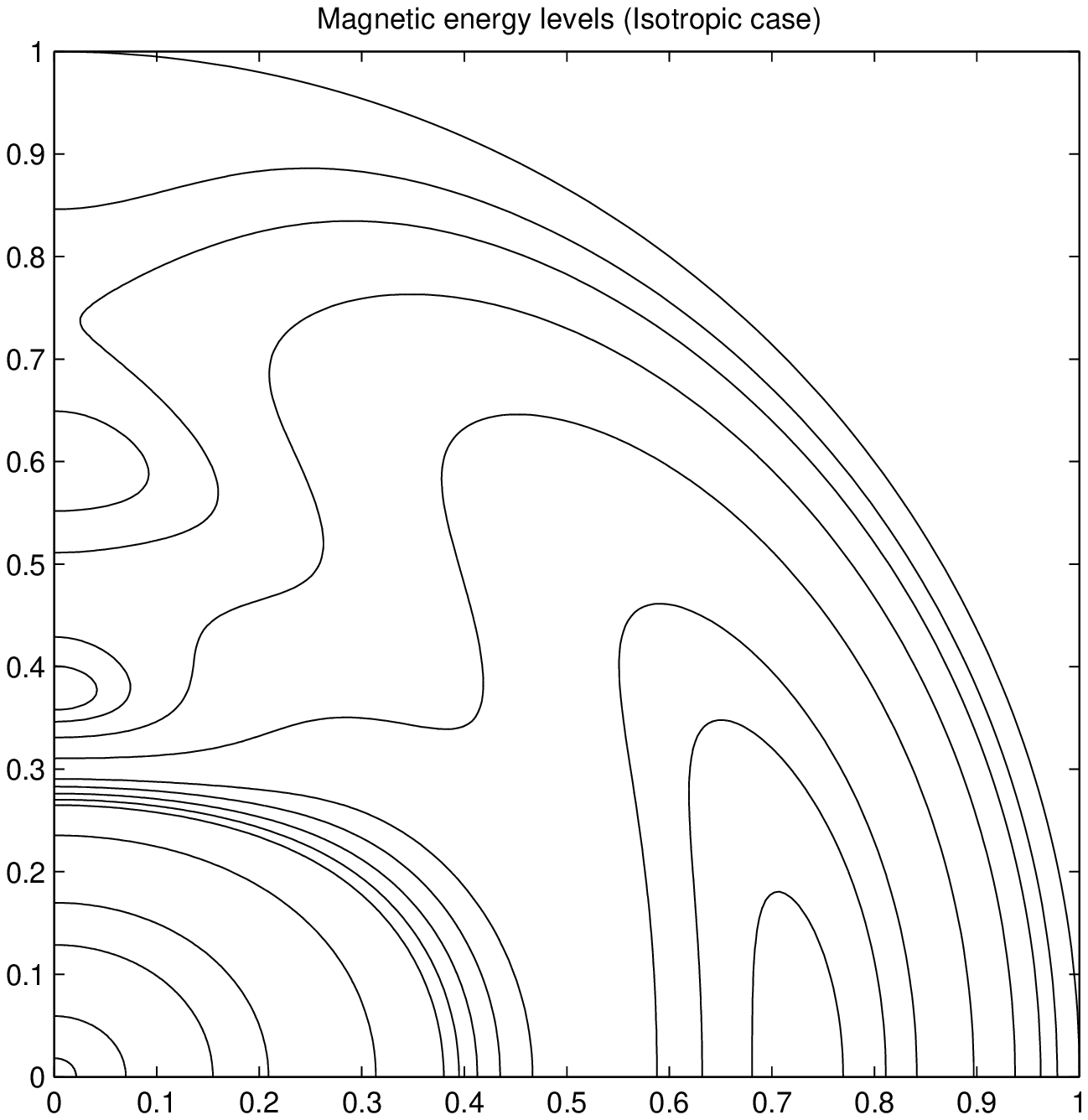}
\includegraphics[width=2.5in,height=2.5in]{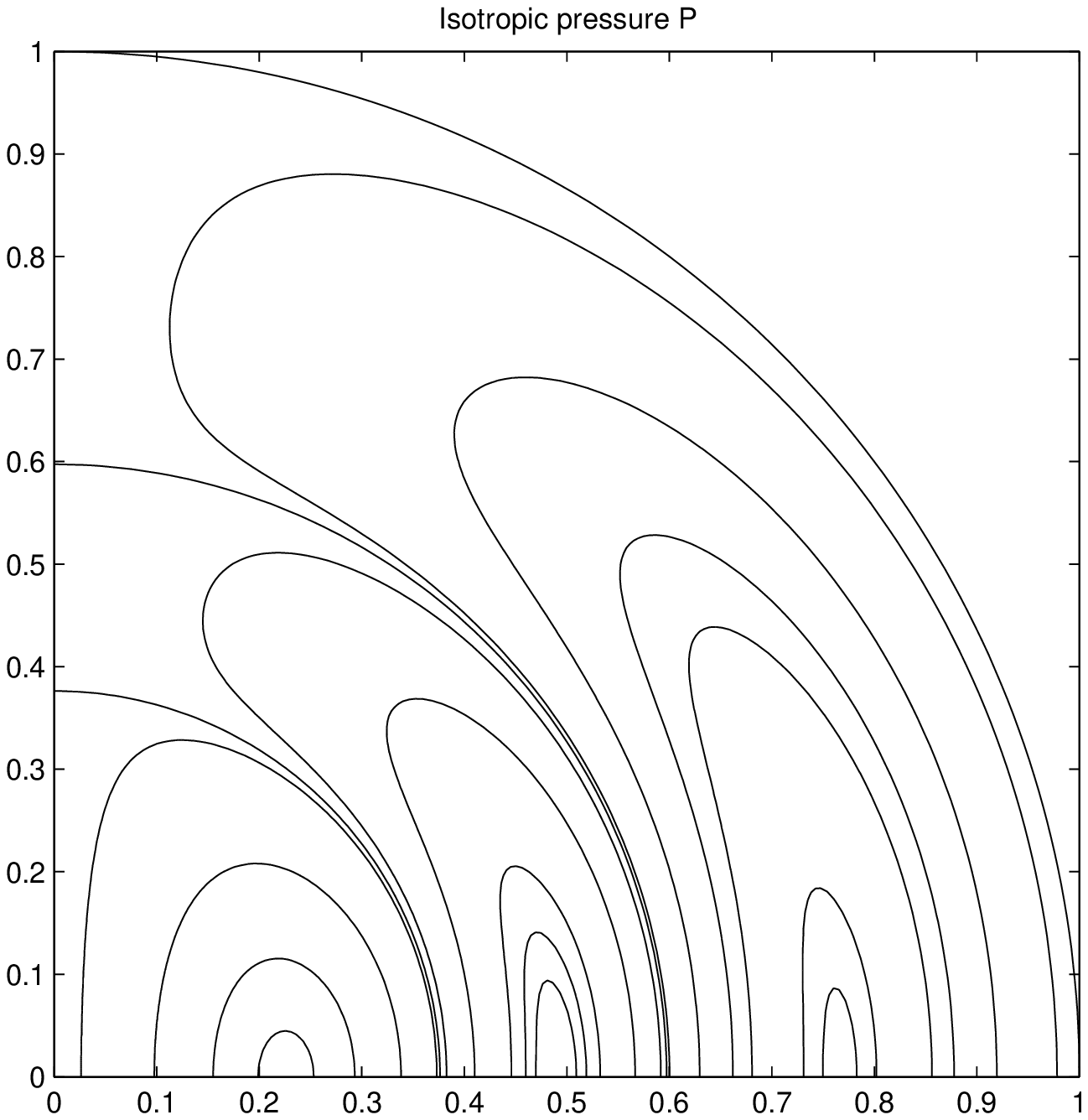}
}

\caption{ \label{fig1_1} Constant levels of magnetic energy density (left) and pressure (right) in the static isotropic MHD
equilibrium (Bobnev's solution (\ref{bob_sol1},\ref{bob_sol2})) for $R=1, \lambda_=\lambda_3 \approx 6.161$. $z$ (upward) is
the symmetry axis, $x$ (horizontal) is the radial axis.}

\end{figure}


\begin{figure}[htbp]
\centerline{
\includegraphics[width=2.5in,height=2.5in]{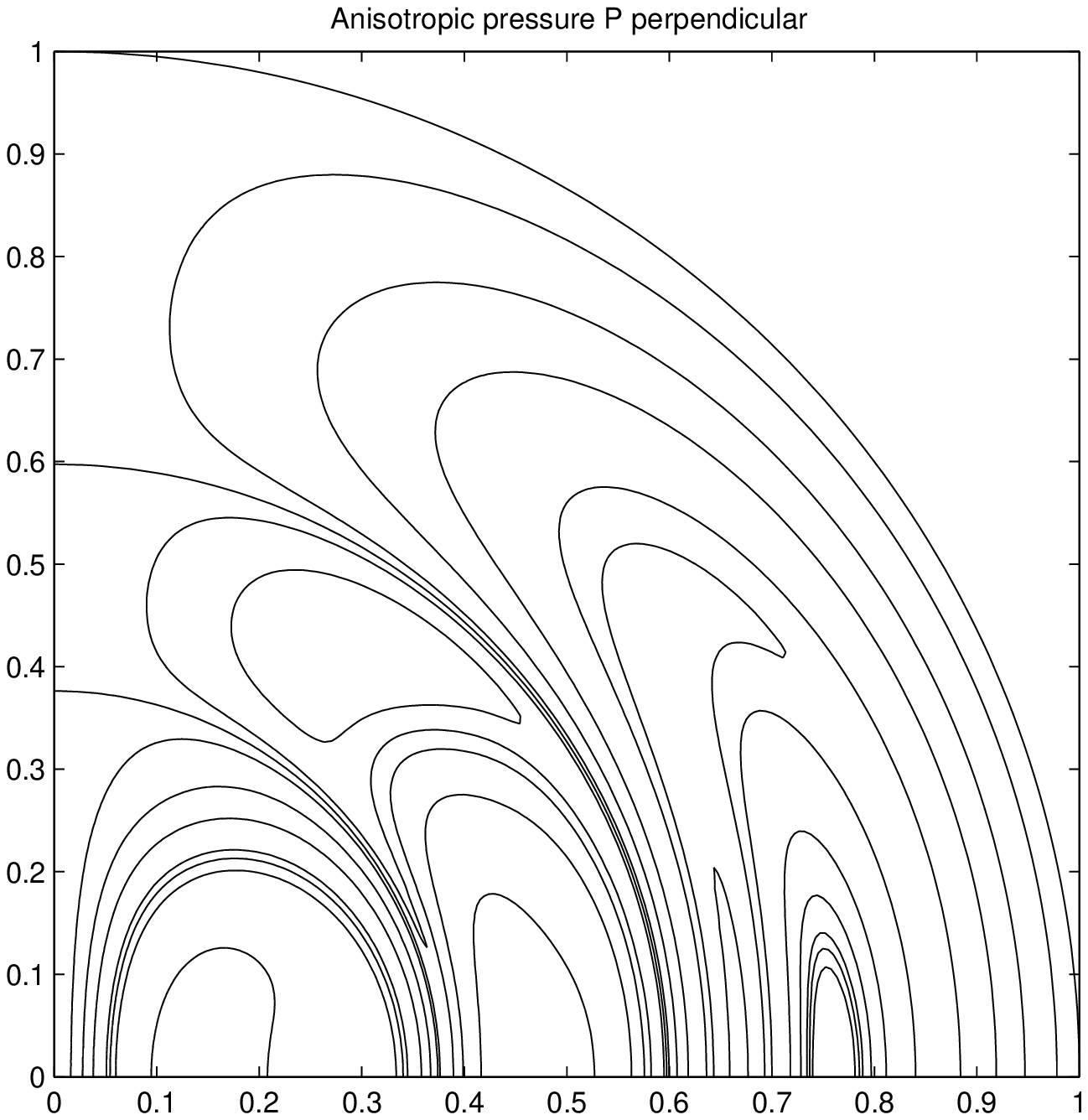}
\includegraphics[width=2.5in,height=2.5in]{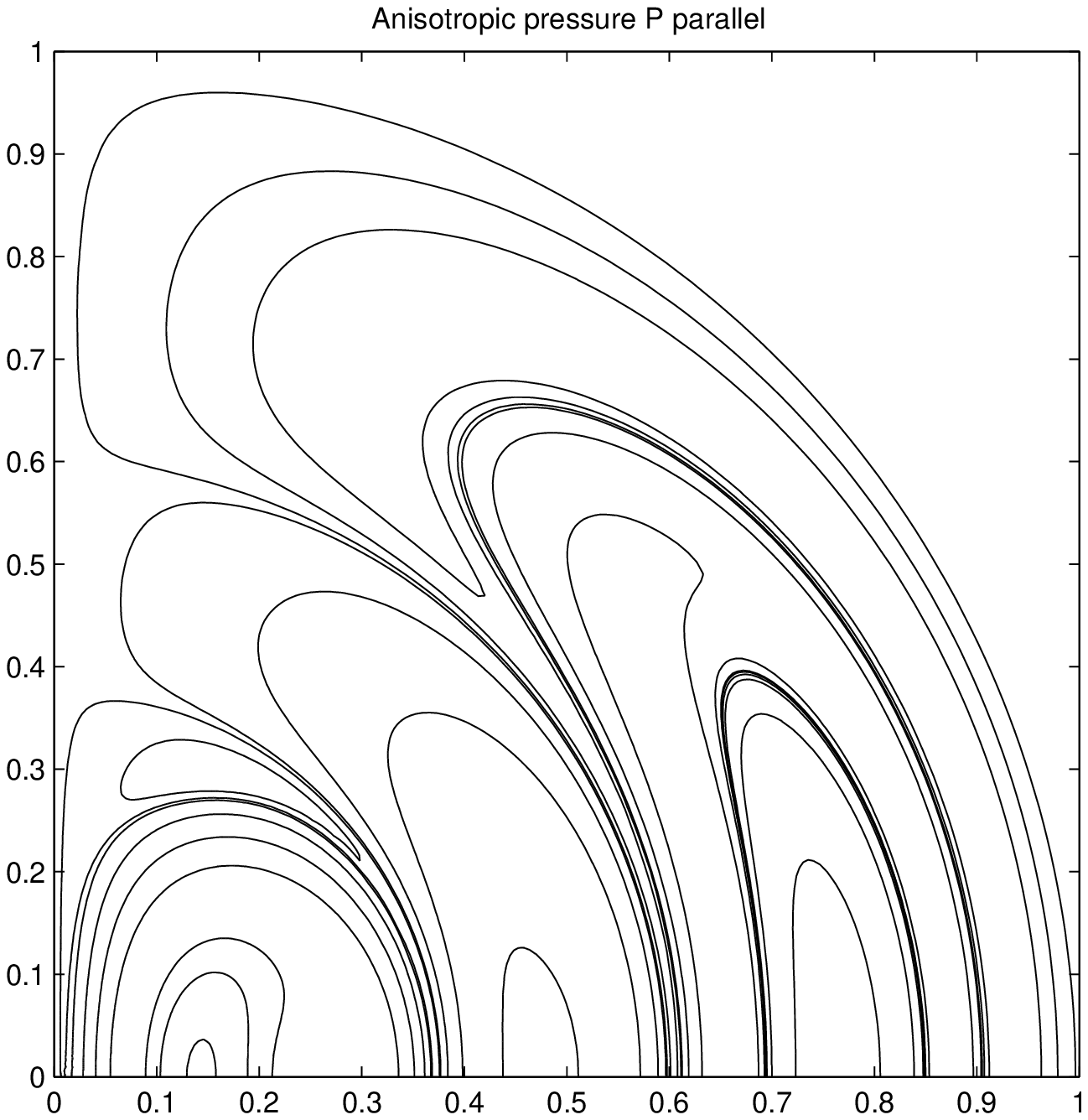}
}

\caption{\label{fig1_2} Surfaces of constant level of plasma pressure components $P,~p_{\perp}~,p_{\parallel}$ in an
anisotropic vortex. }

\end{figure}


\begin{figure}[htbp]
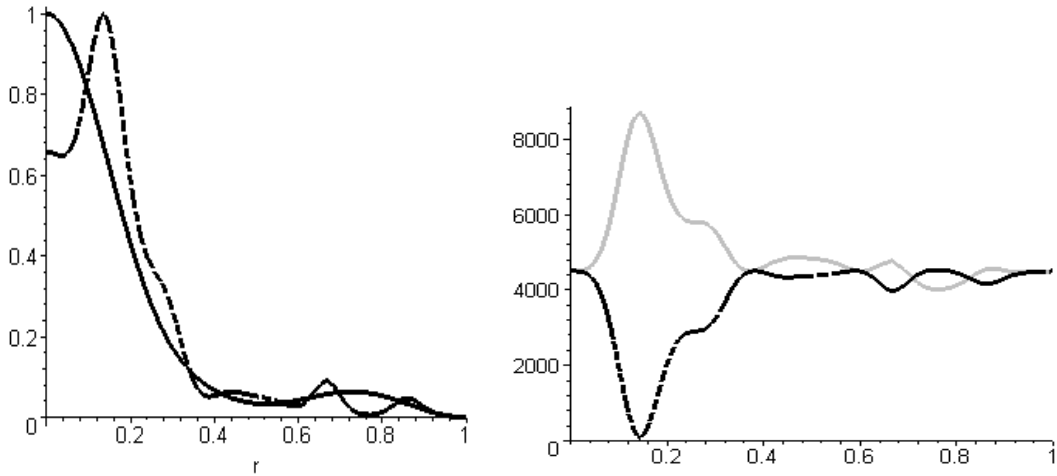

\centerline{
\includegraphics[width=2.5in,height=2.5in]{2_B2_profiles.eps}
\includegraphics[width=3in,height=2in]{2_pr_prof.eps}
}

\caption{\label{fig1_3} Left: Comparison of energy densities ${\bf{B}}_v^2/2$ and ${\bf{B}}_1^2/2$ for isotropic and
anisotropic plasma vortex in the radial direction. Right: Comparison of plasma pressure components
$P,~p_{\perp}~,p_{\parallel}$ in an anisotropic vortex in the radial direction.}

\end{figure}

Similarly to the isotropic vortex, the anisotropic solution has the boundary condition ${\bf B}|_{\rho=R}=0,$ and ${\bf
B}|_{\rho=0}=B_0 {\bf e}_z$ at the origin. The magnetic field is chosen to be zero outside the plasma domain.

As seen from Fig. {\ref{fig1_3}} (right) and expressions (\ref{eq_coroll1_2}),(\ref{eq_coroll1_3}), plasma pressure
components $p_{\parallel 1}, p_{\perp 1}$ of the anisotropic vortex are different within the plasma domain $r<1$, but equal
on the boundary. Thus the presented solution describes a localized anisotropic plasma formation confined by external gas
pressure. (We note that both known vortex-like isotropic solutions, by Bobnev \cite{ball4} and by Kaiser and Lortz
\cite{ball2} have the same feature: the plasma ball is confined by outer gas pressure.)

\section{Further remarks}\label{sec_further}
\smallskip

\noindent\textbf{1.} The Lie symmetry-based approach remains to be one of the most generally applicable and successful
coordinate-independent techniques for generating exact solutions and reduced (invariant) versions of nonlinear partial
differential equations for obtaining self-similar solutions. For an ODE, knowledge of Lie symmetries always leads to the
reduction of order of the equation under consideration, by the number up to the dimension of the symmetry group
\cite{blum_kumei}. The global form of Lie point symmetries is used to generate several-parameter (in some important cases,
infinite-parameter) families of exact solutions from known ones.

In this paper we have presented an example when the complete straightforward Lie analysis reveals an infinite-dimensional
symmetry group for static anisotropic (Chew-Goldberger-Low) Plasma Equilibrium equations. The existence of this symmetry Lie
group leads to a dramatic extension of the set of known solutions, bringing in physically relevant anisotropic static plasma
configurations related to solutions of static MHD equations. In particular, one may construct new static CGL equilibria in
axial and helical symmetry using corresponding Grad-Shafranov and JFKO equations (Section \ref{subs_GS_CGL}), or from known
static MHD configurations (as illustrated by the example in Section \ref{subs_examp_vort}).

\bigskip\noindent\textbf{2.} The potential symmetry theory \cite{blum_kumei} and the algorithmic framework for nonlocally-related
potential systems and subsystems \cite{BC1} has been demonstrated to be useful for calculating new nonlocal symmetries and
new nonlocal conservation laws for a given system of PDEs, especially in the case of two independent variables. However in a
PDE system with $n\geq 3$ independent variables, such as MHD or CGL systems discussed in this paper, the corresponding
potential system is underdetermined, and requires suitable gauge constraints (in the form of additional equations on the
potential variables) to be imposed in order to find nonlocal symmetries \cite{AB_Max}. The question of choosing gauge
constraints that lead to potential symmetries is yet to be answered. Future work will concentrate on this problem, in
particular, for the system of static MHD equations, which can be totally written as a set of four conservation laws
\cite{afc_phd}.

\bigskip\noindent\textbf{3.} Systems of static MHD (\ref{eq_PEE}) and CGL (\ref{eq_APEE})-(\ref{eq_tau_cond}) equations
are examples of systems where the use of appropriate analytical computation software is helpful for the full point symmetry
analysis and leads to finding new symmetry structure (Section \ref{Point_Transfs}.) In the same manner, infinite symmetries
for dynamic MHD \cite{obsymm} and CGL \cite{afc_ob} equilibrium equations can be found within Lie group analysis framework
\cite{afc_pla,afc_kiev}, but have been overlooked due to incomplete analysis in the preceding literature (e.g. \cite{ibr}).

The routines of \verb"GeM" package for Maple presented and used in this work (and in parts in \cite{obsymm, afc_kiev})
perform local (Lie, contact, Lie-B\"{a}cklund) and nonlocal symmetry and conservation law analysis of ordinary and partial
differential equations automatically, i.e. without human intervention. Corresponding determining systems are automatically
generated and reduced, and in most cases completely solved. The package routines allow the solution of important
symmetry/conservation law classification problems for systems with arbitrary constitutive functions. To the best of the
author's knowledge, \verb"GeM" is the most complete and fast end-user-oriented package for Maple.

Further development of the \verb"GeM" package that will include routines for explicit reconstruction of conservation law
fluxes and densities and automatic construction of trees of nonlocally-related potential systems and subsystems \cite{BC1}
is one of the directions of current work.

\bigskip\noindent\textbf{4.} Further research, jointly with G. Bluman and other collaborators, will include the complete
symmetry and conservation law analysis of dynamic CGL and MHD systems, as well as the construction of trees of
nonlocally-related potential systems and subsystems and search for nonlocal symmetries and conservation laws of Nonlinear
Telegraph (NLT) equations and equations of gas/fluid dynamics (which is partly complete).

\bigskip \noindent \textbf{Acknowledgements}

The author is grateful to George Bluman and Oleg I. Bogoyavlenskij for the discussion of results.

-----------------------------------

\pagebreak

\pagebreak
\renewcommand{\theequation}{\Alph{section}.\arabic{equation}}

\begin{appendix}

\section{"GeM" Symmetry computation module: a brief routine description}

Here we outline the program for group analysis of static MHD equations (\ref{eq_PEE}) using the \verb"GeM" package.

First, dependent and independent variables, free functions and free constants are declared. For example, for the CGL static
equilibrium system (\ref{eq_APEE}), the statement is
\begin{equation}
\verb"acgen_decl_vars([x,y,z], [B1(x,y,z),B2(x,y,z),B3(x,y,z),P(x,y,z)],[],[],0);"\nonumber
\end{equation}
Here $B1,B2,B3$ are three components of the magnetic field vector $\bf B$; the empty bracket pairs $[~]$ indicate that no
arbitrary functions and constants are present; $0$ stands for the highest order of derivative on which arbitrary functions
depend.

\smallskip
Then equations are declared by calling the routine
\begin{equation} \nonumber.
\begin{array}{ll}
\verb"acgen_decl_eqs( "\\
\verb"[ equ[1], equ[2], equ[3], equ[4] ], "\\
\verb"1,"\\
\verb"[diff(B1(x,y,z),x), diff(P(x,y,z),x), diff(P(x,y,z),y), diff(P(x,y,z),z)]"\\
\verb");"\\
\end{array}
\end{equation}
where \verb"equ[1]", ...,\verb"equ[4]" are properly prepared Maple expressions for static MHD equations (\ref{eq_PEE}).

Within this routine, all derivatives are defined as symbols: $\partial B_1 /
\partial x\equiv \verb"B1x"$, etc. For example, the Maple expression for
the equation ${\rm{div}}~{\bf{B}}= 0$
\begin{equation}
{\frac {\partial }{\partial x}}{\it B1} \left( x,y,z \right) +{\frac {
\partial }{\partial y}}{\it B2} \left( x,y,z \right) +{\frac {
\partial }{\partial z}}{\it B3} \left( x,y,z \right) = 0 \nonumber
\end{equation}
is represented as
\begin{equation}
\verb"B1x + B2y + B3z = 0"\nonumber.
\end{equation}

After this conversion, the unknown tangent vector field components $\xi^i, \eta^k$ are not composite functions but functions
of all "independent variables", which include dependent and independent variables of the system under consideration and all
partial derivatives.

\medskip
The variables on which tangent vector field components $\xi^i, \eta^k$ depend are specified by using
\begin{equation}
\verb"acgen_def_vf([B1(x,y,z),B2(x,y,z),B3(x,y,z),T(x,y,z),P(x,y,z)],0,0);"\nonumber
\end{equation}

The function names of resulting tangent vector field components are stored in a set \verb"V_F_COORDS_".

\medskip
The overdetermined system of determining equations system variables on which depend are determined by using
\begin{equation}
\verb"acgen_get_split_sys();"\nonumber
\end{equation}

(which generates 133 equations) and reduced by
\begin{equation}
\verb"rss:=rifsimp(sys_l2,convert(V_F_COORDS_,list));"\nonumber
\end{equation}
to obtain 49 simple equations found in the variable \verb"rss[Solved]".

The final answer for tangent vector field components in this case is easily obtained by hand, or Maple built-in integrator
\begin{equation}
\verb"expand(pdsolve(rss[Solved]));"\nonumber
\end{equation}
and provides the results listed in the formulation of Theorem \ref{th_01}.

We remark again that the built-in Maple routine "pdsolve" for PDE integration sometimes produces incomplete answers and
therefore is recommended only for preliminary analysis; the completeness of its output is to be verified by hand in every
case.

\section{Proof of Theorem 2.}\label{sec_ap_proofs}

\bigskip \noindent \textbf{Proof of Theorem 2.}

First we prove the formula (\ref{eq_APEE2_1}). Using the vector calculus identity
\begin{equation}\nonumber
\curl(a{\bf s}) = a \curl({\bf s}) + (\grad a) \times {\bf s}
\end{equation}
and the formilas ${\bf{B}}\cdot\grad\tau=0$ and (\ref{eq_APEE}), one gets
\begin{equation}\nonumber
\curl\left(\sqrt{1-\tau}{\bf{B}}\right)  \times \left(\sqrt{1-\tau}{\bf{B}}\right) = (1-\tau)\curl {\bf{B}} \times {\bf{B}}
- \frac{1}{2}\frac{\sqrt{1-\tau}}{\sqrt{1-\tau}} (\grad \tau \times {\bf{B}})\times {\bf{B}}
\end{equation}
\begin{equation}\nonumber
 = {\rm{grad}}~p_\perp +\tau~ {\rm{grad}}\frac{{{\bf{B}}^2 }}{2} + \frac{{\bf{B}}^2}{2} \grad \tau = \grad\left(p_\perp +\tau\frac{{{\bf{B}}^2
 }}{2}\right).
\end{equation}

We thus observe that the quantity $p_\perp+\tau \frac{{\bf{B}}^2}{2}$ is constant on magnetic field lines:
$\grad\left(p_\perp +\tau\frac{{{\bf{B}}^2 }}{2}\right) \perp {\bf{B}}$. The same is true for $\tau$. Hence without loss of
generality
\begin{equation}\nonumber
F\left(\tau, p_\perp + \tau \frac{{\bf{B}}^2}{2}\right) = F(\Psi),
\end{equation}
where $\Psi$ is a function constant on magnetic field lines.

We may also check that the quantity $p_\perp+\tau \frac{{\bf{B}}^2}{2}$ is invariant under the action of $X_{\infty}$:
\begin{equation}
\frac{\partial}{\partial \varepsilon}\left(p_\perp+\tau \frac{{\bf{B}}^2}{2}\right)'|_{\varepsilon=0}=
\frac{\partial}{\partial \varepsilon}p'_{\perp}|_{\varepsilon=0} + \frac{{\bf{B}}^2}{2}\frac{\partial}{\partial
\varepsilon}\tau'|_{\varepsilon=0}  + \tau\frac{\partial}{\partial \varepsilon}\frac{{{\bf{B}'}^2}}{2}|_{\varepsilon=0}=0;
\nonumber
\end{equation}
hence $\Psi$ may be chosen so that it is also invariant under the action of $X_{\infty}$.

\smallskip
This allows the integration of Lie equations (\ref{e27_5}), and the global group of transformations is
\begin{equation}
B_{k, 1}  = B_k \exp\left(a F\left(\Psi\right)\right),~~ (1-\tau_1)=(1-\tau)\exp\left(-2\varepsilon
F\left(\Psi\right)\right), \nonumber
\end{equation}
\begin{equation}
p_{\perp 1}= p_{\perp} + \frac{{\bf{B}}^2}{2}\left(1-\exp\left(-2\varepsilon F\left(\Psi\right)\right)\right), \nonumber
\end{equation}

In (\ref{eq_coroll1_1}),(\ref{eq_coroll1_2}), $M$ is a sufficiently smooth function separated from zero. It is connected
with F as follows. The case $M\geq0$ is obtained by assigning
\begin{equation}
M(\Psi):=\exp\left(\varepsilon F\left(\Psi\right)\right), \nonumber
\end{equation}
where F is an arbitrary smooth real-valued function; the case $M<0$ - the same way by taking
\begin{equation}
F\left(\Psi\right) = \frac{i\pi}{\varepsilon}+ G\left(\Psi\right), \nonumber
\end{equation}
where $G$ is an arbitrary smooth real-valued function.

This completes the proof of Theorem 2.$\blacksquare$

\end{appendix}


\begin{thebibliography}{9}
\bibitem{RifDesc} The packages for Maple such as \verb"Rif" (the extended version
of the "Standard Form" package by Reid and Wittkopf  and \verb"diffgrob2" are frequently used for the purpose of reduction
of large PDE systems.
\end{thebibliography}
\end{document}